\pdfoutput=1
\documentclass[%
reprint,
 amsmath,amssymb,
 aps,
floatfix,
]{revtex4-1}
\usepackage[activate={true,nocompatibility},final]{microtype}

\usepackage{graphicx}
\usepackage{dcolumn}
\usepackage{bm}
\usepackage{braket}
\usepackage{units}

\newcommand{\aff}{\ensuremath{\mathcal{A}}}
\newcommand{\bindEn}{\ensuremath{\Delta E}}
\newcommand{\freeEn}{\ensuremath{\Delta \mathcal{F}}}
\newcommand{\bigO}{\ensuremath{\mathcal{O}}}
\renewcommand{\vec}{\mathbf}
\newcommand{\tiltOp}{\ensuremath{\mathcal{L}}}
\newcommand{\vecStatDist}{\vec{p}^{\mathrm{s}} }
\newcommand{\statDist}{p^{\mathrm{s}}}
\newcommand{\xinit}{\ensuremath{x_0}}
\newcommand{\wpOne}{w_{1+}}
\newcommand{\wmOne}{w_{1-}}
\newcommand{\wpTwo}{w_{2+}}
\newcommand{\wmTwo}{w_{2-}}
\newcommand{\won}{w_\mathrm{on}}
\newcommand{\woff}{w_\mathrm{off}}
\newcommand{\eul}{\mathrm{e}}
\newcommand{\pbound}{p_{\mathrm{on}}^{\mathrm{s}}}
\newcommand{\punbound}{p_{\mathrm{off}}^{\mathrm{s}}}
\newcommand{\Emin}{\Delta E_{\mathrm{crit}}^{\mathrm{min}}}
\newcommand{\Emax}{\Delta E_{\mathrm{crit}}^{\mathrm{max}}}
\newcommand{\arcoth}{\ensuremath{\mathop{\mathrm{arcoth}}}}
\newcommand{\bareDiff}{\mathcal{D}}
\newcommand{\cum}{\mathcal{C}}
\newcommand{\diff}{\mathrm{d}}
\newcommand{\kb}{k_{\mathrm{B}}}
\newcommand{\vaarw}{v}
\newcommand{\Daarw}{D}

\begin{document}
\title{Force-dependent diffusion coefficient of molecular Brownian ratchets}
\author{Matthias Uhl and Udo Seifert}
\affiliation{II. Institut f\"ur Theoretische Physik, Universit\"at Stuttgart,
70550 Stuttgart, Germany}
\date{\today}
\begin{abstract}
	We study the mean velocity and diffusion constant in three related models
	of molecular Brownian ratchets. Brownian ratchets can be used to
	describe translocation of biopolymers like DNA through
	nanopores in cells in the presence of chaperones on the trans side of the
	pore. Chaperones can bind to the polymer and prevent it from sliding back
	through the pore.
	First, we study a simple model that describes the translocation in terms of
	an asymmetric random walk. It serves as an introductory example but
	already captures the main features of a Brownian ratchet.
	We then provide an analytical expression for the diffusion constant in the
	classical model of a translocation ratchet that was first proposed by
	Peskin et al.\,. This model is based on the assumption that
	the binding and unbinding of the chaperones are much faster than
	the diffusion of the DNA strand. To remedy this shortcoming, we propose a
	modified model that is also applicable if the (un)binding rates are finite.
	We calculate the force-dependent mean velocity and diffusivity for this
	model and compare the results to the original one. Our analysis shows that
	for large pulling forces the predictions of both models can differ strongly
	even if the (un)binding rates are large in comparison to the diffusion
	timescale but still finite.  Furthermore, implications of the
	thermodynamic uncertainty relation on the efficiency of Brownian ratchets
	are discussed.
\end{abstract}

\maketitle

\section{Introduction}
\label{sec:introduction}

Translocation of polymers through nanopores lies at the heart of many processes
in biology.  How exactly do polymers like single stranded DNA or a
RNA chain pass through protein channels embedded in the membranes of the
cell nucleus or the cell proper? This question is of interest both in
biology
and in soft matter physics~\cite{simon_what_1992, zandi_what_2003,
lim_mechanical_2006, chen_ins_2005,palyulin_polymer_2014,
neupert_perspective_2015,knyazev_driving_2018}.

There are several different mechanisms that have been identified to help
sliding nanoscopic chains through pores in an environment with thermal
fluctuations.
If the chain is charged, an electric field in the vicinity of the
pore can pull the chain from the cis side to the trans side. Incidentally,
measuring certain ionic currents that flow through the pore if the pore is
clogged by a monomer can be used to identify that monomer.
Artificially designed nanopores could therefore make it possible to sequence
DNA as it passes through the pore, potentially outperforming other
methods~\cite{bezrukov_counting_1994, kasianowicz_characterization_1996,
fyta_translocation_2011}.

There are also entropic forces at play in the translocation process since the
possible configurations of the chain are restricted by the pore.
Flexible polymer chains can, for example, retract from the pore after they have
already partially translocated if the translocated segment straightens and
enters a hairpinlike state~\cite{fyta_hydrodynamic_2008}.

Yet a different mechanism that aids the translocation, which is the
main interest of this paper, is provided by the binding of certain molecules to the
chain on the trans side of the pore. In the context of protein translocation
this role can be played by chaperone molecules, which are also
important in the context of protein folding and can bind to the chain. Since
chaperones are too large to pass through the pore, the movement of the chain
gets biases towards the side with the higher concentration of chaperones.
Translocation ratchets have been discussed in the context of various different
translocation processes in biological systems~\cite{hepp_bacterial_2017, craig_hsp70_2018}.
Recent experiments showed that the uptake of DNA molecules by Neiseria
gonorrhoeae bacteria~\cite{hepp_kinetics_2016} can be described by a rather
simple translocation ratchet model~\cite{peskin_cellular_1993}. In this case
ComE proteins in the periplasm can bind to the DNA chain and act in the same
way as chaperones do in the case of protein translocation by preventing the
strand from sliding out if they are bound.

Usually theoretical models predict only the mean velocity or the mean translocation
time. It can, however, also be of interest to compare predictions of the
diffusion constant, i.\,e.\,, the rate at which the variance of the traveled
distance increases, from Brownian ratchet models with the experimental findings.
Apart from studies like \cite{krapivsky_fluctuations_2010} predictions for the
diffusion coefficient are sparse. For this reason, experimental data can only
be compared to predictions of the mean velocity. Thereby, a significant part of
information contained in probability distributions gained in experiments does
not enter into fits of the model. The present paper aims to close this gap.

Considering the diffusion constant can also be of theoretical interest as
recent developments regarding the thermodynamic uncertainty
relation~\cite{barato_thermodynamic_2015, gingrich_dissipation_2016, pietzonka_finite-time_2017a, horowitz_proof_2017a} have shown. This relation
shows that high precision, i.\,e.\,, small fluctuations of
quantities like performed work or traveled distance comes at the cost of a high
dissipation rate. Since the uncertainty relation requires only a few
assumptions, which are generically valid for processes at fixed temperature, it
implicates a quite general relation between diffusion coefficient and mean
velocity also for biological processes like translocation.

Covering the translocation process in its full complexity requires the use of
involved numerical simulations. In the case of protein translocation with
Hsp70, for example, chaperones models have been proposed that include forces
applied by the chaperones to the strand, actively pulling it
in~\cite{delosrios_hsp70_2006}. We will, however, focus on the ratcheting
mechanism that is caused by the binding of chaperones to the chain.
Established models usually assume that the parts of the strand on both sides
are in chemical equilibrium with their respective environments, which allows
for the identification of a free energy associated with the position of the
strand~\cite{ambjornsson_chaperone-assisted_2004, ambjornsson_directed_2005, abdolvahab_analytical_2008,
metzler_polymer_2010, abdolvahab_sequence_2011,
abdolvahab_chaperone-driven_2017}.
In contrast, we use a more detailed description in the spirit of the classical
model of a translocation ratchet introduced by Peskin
et\,al.~\cite{peskin_cellular_1993}. The basic concept of our models is that
the movement of the DNA strand would be governed by a process without
directional preference if it were in equilibrium.
However, on the trans side of
the pore molecules bind to equidistant binding sites on the strand with rate
$\won$ and unbind with rate $\woff$. An occupied binding site can not pass
through the pore. As a result the movement of the DNA is biased and can be used
to perform work against a force.
We assume that the strand is stiff. Effects of bending of the chain are
investigated, e.\,g.\,, in~\cite{yu_polymer_2014a, adhikari_translocation_2015, suhonen_chaperone-assisted_2016, mondal_ratchet_2016, emamyari_polymer_2017}.
Furthermore, we neglect the finite length of
the strand and assume that the movement can go on in both directions
indefinitely.

The paper is organized as follows. In
Sec.~\ref{sec:alternating_asymmetric_random_walk}, we introduce a simple
model for rectification based on an asymmetric random walk with an alternating
set of rates. Although simple, the model shows two distinct features that we
encounter also in more involved models. The strand is drawn in, when it is not
pulled outwards by an external force. If it is, however, pulled strongly, the
mean velocity reaches a finite limit.
The classical model for rectified Brownian motion that was introduced in
\cite{peskin_cellular_1993} is briefly recapitulated in
Sec.~\ref{sec:oster_model}. In addition to the mean velocity, which was
already calculated in the original work, we provide an analytical expression
for the diffusion coefficient for this model.
The key assumption that lets us treat the classical model analytically is the
time-scale separation between the diffusion process of the strand and the
binding process of the molecules that block transitions through the pore. A
 model that drops this assumption is studied numerically in
Sec.~\ref{sec:oster_model_with_memory}.
As a consequence of the thermodynamic uncertainty relation, which connects mean
velocity, diffusion coefficient, and entropy production, the efficiency of
the ratcheting process is bounded by a function that only depends on the
measurable quantities $v$, $D$, and $f$. We study such thermodynamic
implications provided by the uncertainty relation in
Sec.~\ref{ssec:thermodynamics}.  We summarize our findings in
Sec.~\ref{sec:conclusion}.

\vspace{-0.2cm}
\section{Simple Model}
\label{sec:alternating_asymmetric_random_walk}
\vspace{-0.1cm}

\subsection{States and rates}
\label{sub:definition}
\vspace{-0.1cm}

\begin{figure}
	\centering
	\includegraphics[scale=1]{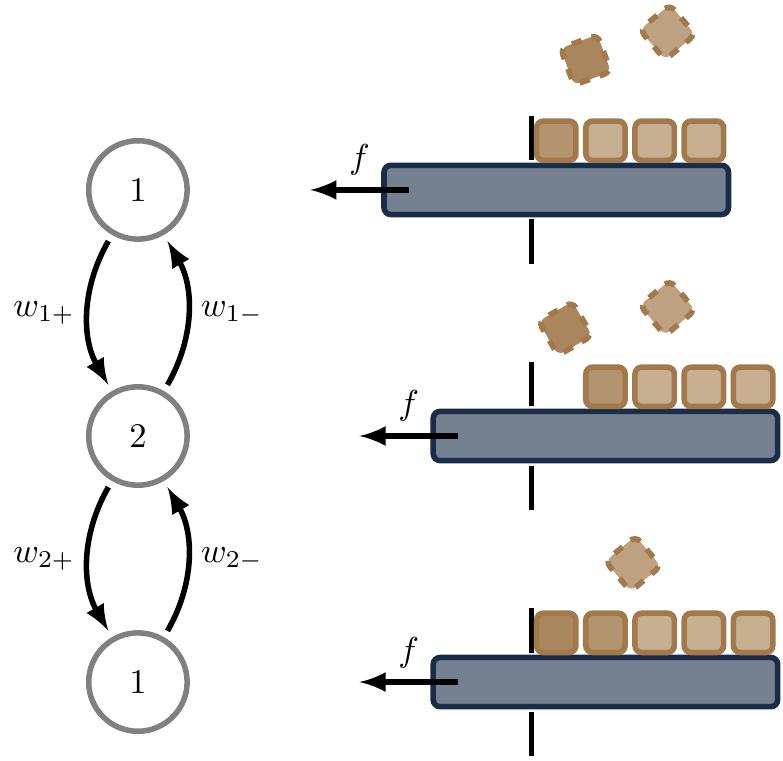}
	\caption{Schematic depiction of the alternating random walk model. The
	system can be in two different sates depending on whether the binding site
in front of the pore is occupied or not. The state can either change by binding
or unbinding of the chaperone on this site or by movement of the hole strand.
These transition rates are assumed to obey a local detailed balance condition
relating these rates to the binding energy and the pulling force.}
	\label{fig:simple_model}
\end{figure}

The assumptions entering this model are that the probability of finding two
consecutive empty binding sites is negligibly small and that it is possible to
map the movement of the chain through the pore to a discrete jump process.
Under these conditions the translocation process can be described as a two
state Markov process (see Fig.~\ref{fig:simple_model}). In state 1 all sites on
the trans side are occupied and in state 2 the first site next to the pore is
empty. There are two different ways to get from state 1 to 2. Either the whole
strand moves further to the trans side by one position with rate $\wpOne$,
leaving the first pore empty, or the first chaperone unbinds with rate
$\wmTwo$.  Transitions in the opposite direction are also possible by the same
mechanisms.  The strand can jump backwards with rate $\wmOne$ or the first
binding site gets occupied with rate $\wpTwo$.
Since this simple model is, by design, restricted to situations where it is
unlikely to find two consecutive empty binding sites, the binding rate $\wpTwo$
should always be greater than the unbinding rate $\wmTwo$.

Since the strand is pulled back by a force, the
movement by one position in the forward direction is associated with the free
energy difference $\freeEn = -f$ and the local detailed balance
condition reads
\begin{equation}
	\frac{\wpOne}{\wmOne} = \eul^{-f}\,.
	\label{eq:local_detailed_balance_1}
\end{equation}
Note that we set $\kb T=1$ throughout and use the distance of two neighboring
binding sites as the unit of length, rendering the force $f$ dimensionless.
With this convention both the mean velocity and the diffusion constant have the
unit of a frequency.

We also assume local detailed balance for the (un)binding transition relating
the associated rates with the binding energy $\bindEn$ by the condition
\begin{equation}
	\frac{\wpTwo}{\wmTwo}=\eul^{\bindEn}.
	\label{eq:local_detailed_balance_2}
\end{equation}

Since our goal is to study the force dependence of mean velocity and
diffusivity, we have to specify how each rate depends on the force applied to
the system explicitly. Otherwise we would not be able to compare two
configurations that only differ in the applied force.
Equations~\eqref{eq:local_detailed_balance_1}
and~\eqref{eq:local_detailed_balance_2}, however, only fix the ratio of the
rates in terms of the force and the binding energy.
To fix each rate individually, additional assumptions on the force dependence
have to be made. For the movement of the strand that is modeled by the rates
$\wpOne$ and $\wmOne$ we assume that the pulling force affects the rates in a
symmetrical manner, since the strand should not have a directional preference.
This leads to the ansatz
\begin{equation}
	\wpOne = k\eul^{-f/2} \quad \text{and} \quad \wmOne = k \eul^{f/2}\,,
	\label{eq:load_sharing_f}
\end{equation}
where we introduce the rate constant $k$ that fixes the timescale of the
movement of the strand.
For the dynamics of the (un)binding process of the chaperone molecules there does
not need to exist such a symmetry. We will, however, assume that the binding
and unbinding rates do not depend on the force $f$. Therefore, we make the
general ansatz
\begin{equation}
	\wpTwo = w \quad \text{and}%
	\quad \wmTwo = w \,\eul^{-\bindEn}
\end{equation}
introducing a positive constant $w$. It should be mentioned
that the choice of putting the exponential term from
Eq.~\eqref{eq:local_detailed_balance_2} exclusively into the unbinding rate
$\wmTwo$ has no physical meaning; it is done only because it simplifies some
calculations later on. Any other choice of splitting the exponential term
between the binding and unbinding rate can easily be accommodated by
redefining $w$.

\subsection{Mean velocity and diffusivity}
\label{sub:mean_velocity_and_diffusivity}
In this section we calculate the mean velocity and diffusion coefficient for
the minimal model introduced above. This is done by calculating the rescaled
cumulant generating function (from here on simply called the generating function)
\begin{equation}
	\alpha(\lambda) \equiv \lim_{t \rightarrow \infty}  \frac{1}{t}
		 \ln \left\langle \eul^{\lambda \Delta x}  \right\rangle\,,
\end{equation}
where $\Delta x$ is the distance traveled by the strand during time $t$. The
scaled cumulants $\mathcal{C}_{n} \equiv \lim_{t \rightarrow \infty} C_{n}/t$
are encoded in the generating function as its $n$th derivative
at $\lambda = 0$. The first two cumulants are the expectation value and the
variance of the distribution of $\Delta x$, i.\,e.\,, $C_{1} = \langle \Delta x
\rangle$ and $C_{2} = \langle ( \Delta x - \langle \Delta x \rangle  )^2
\rangle$.
It can be shown that the generating function is the largest eigenvalue of the
so called tilted operator~\cite{lebowitz_gallavotti-cohen-type_1999}.

The tilted operator associated with the total distance traveled by the strand
is given by the matrix
\begin{equation}
	\tiltOp(\lambda) =\left(\! \begin{array}{cc}
			-\wpOne   - \wmTwo & \wmOne \eul^{-\lambda/2}
				+ \wpTwo \eul^{\lambda/2} \\
			\wpOne \eul^{\lambda/2} + \wmTwo \eul^{-\lambda/2}  &
			-\wmOne  - \wpTwo
	\end{array} \!\right).
\end{equation}
It can be obtained from the master operator by multiplying every transition
rate on the off-diagonals with an exponential term of the form $\eul^{d
\lambda}$, where $d$ is the increment of $\Delta x$ for a jump associated
with the specific rate (see, e.\,g.,~\cite{koza_general_1999} for a derivation).

In this particular case the rescaled cumulant generating function can be
obtained directly in closed form as the largest eigenvalue of
$\tiltOp(\lambda)$.
The mean velocity and variance of the position are the first and second
derivatives of this function at $\lambda=0$. The calculations are straight
forward but lead to rather lengthy expressions and are therefore not presented
here. The result for the mean velocity is given by
\begin{equation}
	\vaarw = \frac{\wpOne \wpTwo - \wmOne \wmTwo}{\wpOne + \wmOne +
	\wpTwo + \wmTwo} \,.
\end{equation}
The expression for the diffusion coefficient can be put in the form
\begin{equation}
	\Daarw =\frac{1}{2}  \frac{\wpOne \wpTwo + \wmOne \wmTwo - 2
	\vaarw^{2}  }{\wpOne + \wmOne + \wpTwo + \wmTwo}  \,.
\end{equation}

\subsection{Change of parametrization}

The ansatz for the transitions rates introduces the two timescales $w$
and $k$ that parametrize the local detailed balance condition. These parameters
are, however, not easy to determine experimentally. For this reason we express
$w$ and $k$ in terms of more accessible quantities, namely, the velocity and
diffusion coefficient without pulling force, which we denote by $v_{0}$ and
$D_{0}$, respectively.

Replacing the timescales by velocity and diffusion without force eliminates the
need to know the load sharing entering the binding rates as long as we assume
that it does not depend on the force.
For the jump rates of the strand, $\wpOne$ and $\wmOne$, the load sharing is
assumed according to Eq.~\eqref{eq:load_sharing_f} with constant $k$.

Solving the equations $\vaarw(k,w,\bindEn,f=0) = v_{0} $ and
$\Daarw(k,w,\bindEn,f=0) =D_{0}$ for $k$ and $w$ leads to a quadratic equation
that yields the two sets of solutions
\begin{subequations}
\begin{align}
	k^{^\pm} &= \frac{v_{0}^{2} \left(\eul^{\bindEn } -1 \right) \pm
	v_{0}^{2} \sqrt{C(v_{0} , D_{0}, \bindEn )} }{2 v_{0} \left( \eul^{\bindEn} +1
	\right) - 4D_{0} \left( \eul^{\bindEn}  -1\right)  } \quad
	\text{and}\label{eq:k_sol}  \\
w^{\pm} &= \frac{2 k^{\pm} v_{0} \eul^{\bindEn}}{k^{\pm} \left( \eul^{\bindEn}
-1  \right) - v_{0}\left( \eul^{\bindEn} +1  \right)  } \,,
\end{align}
\label{eq:new_rates}
\end{subequations}
where the discriminant in Eq.~\eqref{eq:k_sol} is given by
\begin{equation}
	C(v_{0} , D_{0}, \bindEn ) \equiv  8\frac{D_{0} }{v_{0} } \left(\eul^{2
	\bindEn} -1   \right) -  3 \eul^{2 \bindEn} -10\eul^{\bindEn} -3  \,.
\end{equation}
With this parametrization one only has to determine which of the two solutions
fits experimental data, instead of finding $k$ and $w$, which leaves
considerably less ambiguity. In the following, we focus mainly on the $+$
solution since in this case both the mean velocity and the diffusion constant
stay finite for any combination of parameters.

\begin{figure*}[tp]
	\centering
	\includegraphics[scale=0.98]{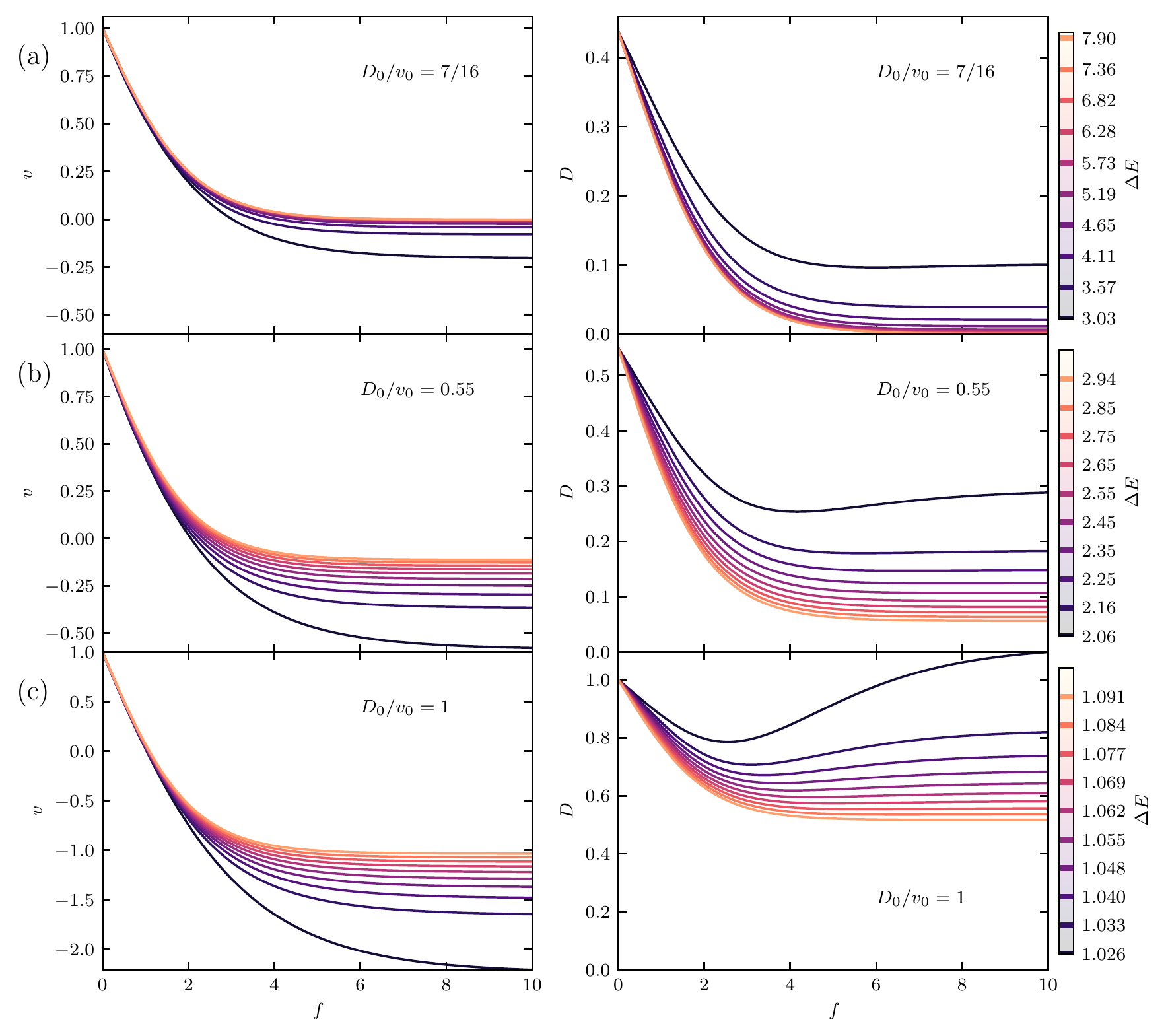}
	\caption{Mean velocity and diffusion coefficient as a function of the
	pulling force $f$ in the simple model. The
	three cases shown represent different ratios of diffusion coefficient and
	mean velocity $D_0/v_0$ in the absence of a pulling force. They correspond
	to the dotted lines depicted in Fig.~\ref{fig:aff_bounds}. The mean
	velocity decreases monotonically in all cases, whereas the diffusion
	coefficient develops a local minima in cases (b) and (c) for binding energies
	close to the lower bound. For large $f$ both quantities reach a stationary
value for large $f$ since the process degenerates to a Poisson process in this
limit.}
	\label{fig:alter_asym_rw_v_D}
\end{figure*}

The rates obtained from Eq.~\eqref{eq:new_rates} are physically meaningful,
i.\,e.\,, $k$ and $w$ are both positive, only if the binding energy is in the
range
\begin{align}
	\bindEn &\geq \Emin \equiv \ln \frac{5 v_{0} + 4	\sqrt{(2D_{0} )^2 + v_{0}^{2}
	} }{8 D_{0} - 3 v_{0}  }
	\label{eq:lower_bound}\\
	\bindEn &< \Emax \equiv \ln \frac{2 D_{0} + v_{0}  }{2D_{0}-v_{0} } =
		2 \arcoth \left( \frac{2 D_{0} }{v_{0} }  \right) \,.
		\label{eq:upper_bound}
\end{align}

At the lower bound the two solutions coincide. Below $\Emin$ the rate constants
would become complex. At the upper bound either $w$ or $k$ diverges. Above
$\Emax$ both rates stay finite but one of the two is negative.

\subsection{Results}
\label{sub:results}

To give an overview of the behavior of the mean velocity and the diffusion
constant as a function of the pulling force $f$,
Fig.~\ref{fig:alter_asym_rw_v_D} shows both quantities obtained using $k^{+}$
and $w^{+}$  for three
different cases of the ratio $D_{0}/v_{0}$.  Here, the color
encodes the binding energy $\bindEn$ of the chaperone molecules.

In case (a) with $D_{0}/v_{0} = 7/16$ there exists no upper limit
to the possible values of $\bindEn$, whereas in case (b) the upper limit exists
but the range of possible values of $\bindEn$ is still fairly wide. Case (c)
refers to parameters such that the upper and lower bound to $\bindEn$ are
close, leaving only a narrow range for the binding energy.
Overall the curves for the mean velocity qualitatively resemble the
experimental findings from~\cite{hepp_kinetics_2016} quite well, despite the
simplicity of the used model.

Remarkably, we find that in case (a), where there is no upper bound to the
binding energy, the behavior of diffusion and velocity does not change
significantly over a comparatively large range of $\bindEn$. Both functions
decrease monotonically with increasing force $f$.

This behavior changes for larger values of $D_{0}/v_{0}$ as it is evident from the
plots for case (b). Here we find a stronger dependence on $\bindEn$. Also, the
diffusion coefficient develops a local minimum as $\bindEn$ approaches the lower
critical value $\Emin$. The mean velocity still decreases monotonically with
rising $f$.

The local minimum of the diffusion coefficient is even more pronounced in case
(c). Here, the bounds of Eq.~\eqref{eq:lower_bound} and
\eqref{eq:upper_bound} restrict $\bindEn$ to a narrow interval. Velocity and
diffusivity are highly sensitive to changes in values of $\bindEn$ within this
interval, as it is shown in the bottom row of Fig.~\ref{fig:alter_asym_rw_v_D}.
This is the case because $k^{+}$ and $w^{-}$ diverge at $\Emax$ and therefore
change drastically within the allowed interval of $\bindEn$.

In all cases, the mean velocity and the diffusion coefficient approach a finite
limit as $f$ increases.  In the limit of very strong pulling, the progression
of the backwards translocation is bottlenecked by the unbinding of the
chaperone molecules. As a result, the process is described by a Poisson process
with the jump rate $\wmTwo = w \eul^{-\bindEn}$ associated to the unbinding.
Consequently, we find the relation between mean value, variance, and jump rate
that is characteristic for such a process
\begin{equation}
	\lim_{f \rightarrow \infty} v = - 2 \lim_{f \rightarrow \infty} D =
	-w \eul^{-\bindEn} \,.
\end{equation}

\subsection{Dynamical phase diagram}
\label{sub:dynamical_phase_diagram}

\begin{figure}
	\centering
	\includegraphics[width=1\linewidth]{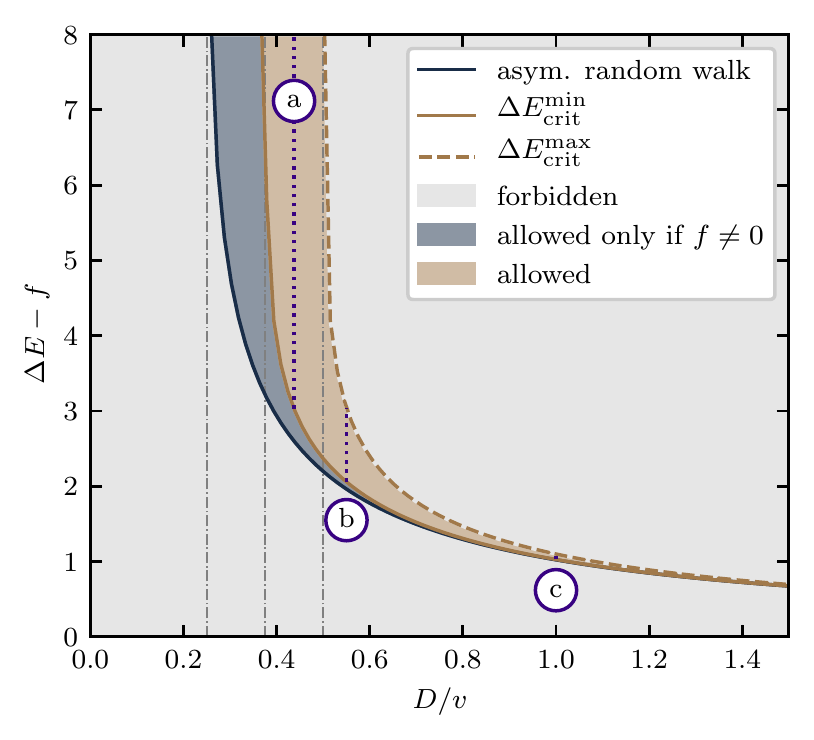}
	\caption{Dynamical phase diagram with bounds on the affinity depending on
		the ratio $D/v$ between diffusion coefficient and mean velocity.
	The region between $\Emin$ and $\Emax$ is accessible for
any value of the pulling force $f$. The region below $\Emin$ can only be reached if the
strand is pulled outwards with $f\neq 0$. A global lower bound on the
affinity is given by the curve generated by an asymmetric random walk. For
vanishing $f$ a stronger bound can be obtained that separates both regions. There is also a global upper bound to the affinity.
The global lower bound diverges at $D/v = 1/4$; the lower bound for $f=0$
diverges at $D_{0}/v_{0}=3/8$  and the global upper bound diverges at
$D/v=1/2$. The three different cases of $D/v$ that are studied in more detail
in Fig.~\ref{fig:alter_asym_rw_v_D} are marked with dotted lines.  }
	\label{fig:aff_bounds}
\end{figure}

The fact that there have to be  bounds on the binding energy under the
constraint of a given value of the velocity and diffusivity could also be
anticipated as a consequence of a more general bound on the diffusion
coefficient introduced in
\cite{barato_thermodynamic_2015, pietzonka_universal_2016}
that bounds the Fano factor, i.\,e.\,, $2 D/v$, through the
affinity $\mathcal{A}$ and the number of states $N$ (in our case two)
\begin{equation}
	\frac{1}{2N} \coth \left( \frac{\aff}{2 N} \right) \leq
	\frac{D}{v} \leq
	\frac{1}{2} \coth \left( \frac{\aff}{2}  \right) \,.
	\label{eq:aff_dep_bounds}
\end{equation}
These bounds are valid for any set of parameters, so they are in particular
applicable to the situation without pulling force considered above.
In this case, we have to replace $v$ and $D$ by $v_{0}$ and $D_{0}$,
respectively.

Without the pulling force, the affinity is equal to the binding energy,
i.\,e.\,, $\aff=\bindEn$.  Obviously, the upper bound in
Eq.~\eqref{eq:aff_dep_bounds} leads to the same upper bound on the binding
energy as in Eq.~\eqref{eq:upper_bound}.

For the lower bound, however, the resulting bound on the binding energy
\begin{equation}
	\Delta E \geq 4 \arcoth \left( \frac{4 D_{0} }{v_{0} }  \right)
\end{equation}
is weaker than the one found in Eq.~\eqref{eq:lower_bound}. This can be
explained by the fact that equality holds in the left inequality in
Eq.~\eqref{eq:aff_dep_bounds} if and only if the process in question is an
asymmetric random walk. Since we set $f=0$, it is not possible to reach an
asymmetric random walk with the remaining parameters without having $v_{0}=0$
and thereby invalidating the bound. The bound in Eq.~\eqref{eq:lower_bound}
represents the state closest to an asymmetric random walk with finite velocity
and $f=0$. This behavior is illustrated in Fig.~\ref{fig:aff_bounds}, where the
accessible region of the affinity is shown depending on the ratio of diffusion
coefficient to mean velocity. The region between $\Emin$ and $\Emax$ contains
the possible values of $\bindEn$ for fixed $D_{0}/v_{0}$.  For arbitrary values
of $f$ the accessible region is larger and restricted by the curve that
corresponds to an asymmetric random walk with the respective ratio $D/v$.
In this diagram, we also mark the three different parameter values that we used
to illustrate the general behavior of the mean velocity and the diffusion
constant in Sec.~\ref{sub:results}.

\section{Classical translocation ratchet}
\label{sec:oster_model}

\begin{figure}
	\centering
	\includegraphics{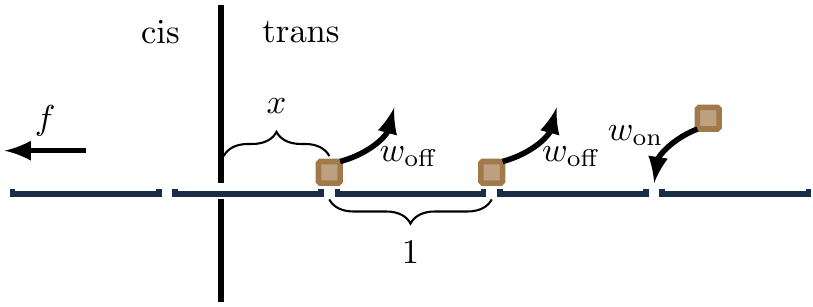}
	\caption{Schematic depiction of a translocation ratchet. The
	position of the strand is encoded in the distance $x$ of the pore from the
closest binding site on the trans side. Each empty binding site gets occupied
with rate $\won$ and each occupied site is emptied with rate $\woff$. }
	\label{fig:setup_cont}
\end{figure}
While the simplistic model of an asymmetric random walk with alternating rates
is able to capture the main feature of the ratcheting mechanism, namely that it
can pull the strand inwards against a force, we will now
focus on a more involved model that should give a better description of the
actual mechanism that governs DNA uptake as reported in
\cite{hepp_kinetics_2016}.

The model presented here was originally introduced
in~\cite{peskin_cellular_1993}, where an expression for the mean velocity of the
ratchet was derived. We augment this calculation by a general scheme by which
not only the mean velocity, i.\,e.\,, the first scaled cumulant of the
distribution of traveled distance, but in fact all cumulants and especially the
diffusion constant can be calculated in a systematic way.

The key idea of this classical model, which is schematically depicted in
Fig.~\ref{fig:setup_cont}, is to encode the position of the strand in a
cyclical variable $x$ that is the distance of the pore to the closest binding
site on the trans side. As before, we use the distance of two neighboring binding sites as
the length scale for the mathematical description. This means that $x$ can take
values in the range $[0,1)$, where the boundaries represent the state where a
new binding site becomes available or vanishes.
It is assumed that the dynamics of the position is described by an overdamped
Brownian motion and it evolves according to the Langevin equation
\begin{equation}
	\dot{x} = -\mu f + \xi(t)\,,
	\label{eq:langevin}
\end{equation}
where $\xi(t)$ denotes a delta correlated white noise with $
\left\langle \xi(t)\xi(t') \right\rangle = 2 \bareDiff \delta(t-t')$ with the
bare diffusivity $\bareDiff$; $f$ is the force
pulling on the strand and $\mu = \bareDiff$ is the mobility.

The state of the binding site  in front of the pore (on the trans side)
determines, whether it is possible for the position to cross the periodic
boundary and retract the ratchet by one step.  Ideally one would have to keep
track of the states of all binding sites and forbid the jump from $x=0$ to $1$
whenever the site in front of the pore is occupied and blocks the transition.
However, this is not necessary in the case of fast (un)binding rates of
the chaperone molecules in which case the binding sites are instantaneously
equilibrated. For this reason it is sufficient to forbid a jump from $x=0$ to
$1$ with the stationary probability $\pbound = \won/(\won + \woff)$  of being
in the bound state. Jumps in the opposite direction are always allowed.

The merit of this, perhaps oversimplified, model lies in the fact that it is
possible to obtain the stationary distribution of the position in closed form.
Based on the stationary distribution it is furthermore possible to determine
the mean velocity of the Brownian ratchet as~\cite{peskin_cellular_1993}
\begin{equation}
	v =  \frac{\bareDiff f^2}{ \frac{\eul^f -1}{1-K(\eul^f -1)} -f
	} \,,
\end{equation}
where $K \equiv {\woff}/{\won}$ denotes the dissociation constant.
As a consequence, the stall force is given by
\begin{equation}
	f_{0} = \ln \left( 1+ \frac{1}{K}  \right)\,.
\end{equation}

While the stationary distribution is sufficient to calculate the mean velocity
as already performed in \cite{peskin_cellular_1993},
the diffusion coefficient, i.\,e.\,, the rate at which the variance of the
position is increasing, cannot be calculated from the distribution alone.
It is, however, interesting to also compare the diffusion coefficient
gained from experimental data with the prediction of this model.
For this reason we present a systematic scheme that allows for the
iterative calculation of mean velocity, diffusion constant and all higher
cumulants.

The calculations are based on the same principles that we used to calculate the
diffusion constant in the simple model. In this case, however, it is to our
knowledge not possible to calculate the largest eigenvalue  $\alpha(\lambda)$
of the tilted operator $\tiltOp(\lambda)$ directly. Since we are only
interested in the first few derivatives of $\alpha(\lambda)$ at $\lambda=0$, it
is possible to expand the largest eigenvalue into a perturbation series and
derive an iteration formula for the cumulants. The details of this expansion
are explained in Appendix~\ref{sec:diffusion_coefficient}.
In the specific case of the Brownian ratchet, the calculation is additionally
complicated by the somewhat unusual form of periodic boundary conditions for
the position variable $x$ that prevent us from directly applying known results
for the tilted operator~\cite{touchette_introduction_2018}.

Nevertheless, it is possible to obtain the tilted operator and perform all
necessary calculations through discretization of the state space. The procedure
is quite technical, which is why it is not presented here. The interested
reader is referred to appendix~\ref{sec:analytical_diffusion_oster}. Here, we
only present the final result for the diffusion coefficient in the classical
model of a Brownian ratchet that constitutes one of the main results of this
paper.  The diffusion coefficient is given by

\begin{figure*}
	\centering
	\includegraphics[scale=1]{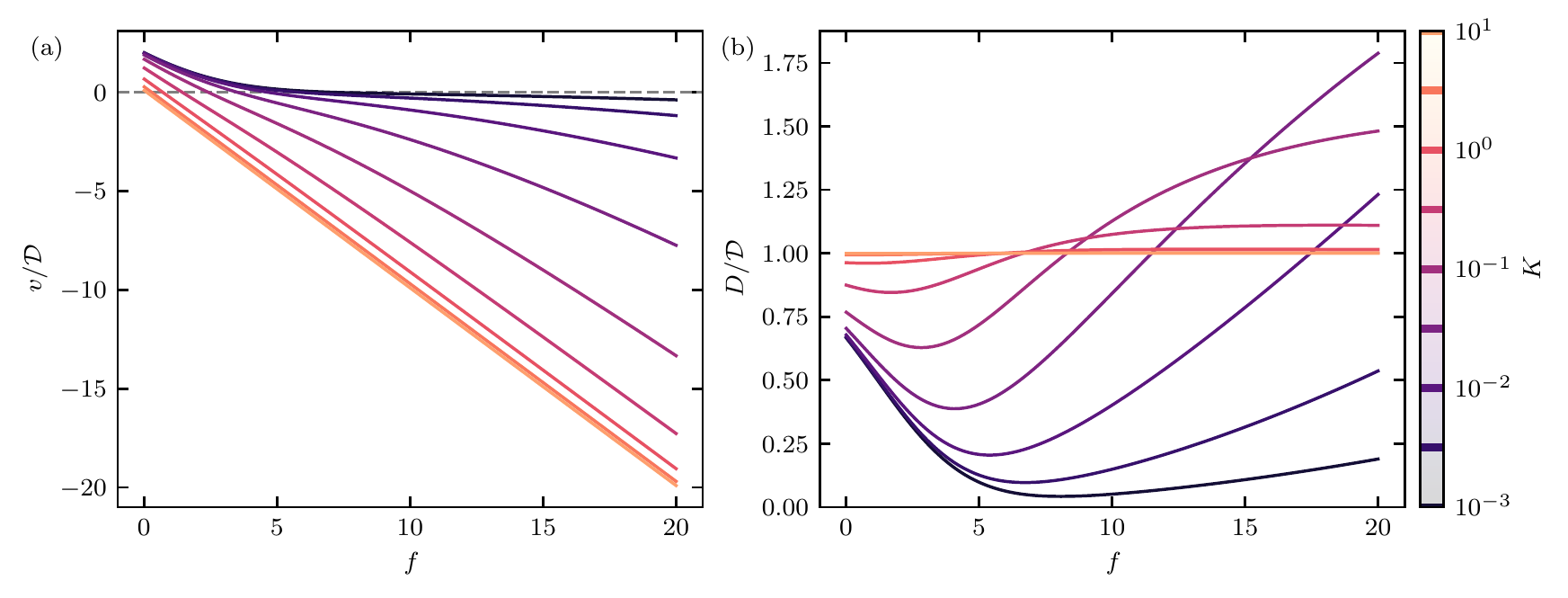}
	\caption{Mean velocity (a) and diffusivity (b) of the classical Brownian
	ratchet. For large $K$, mean velocity and diffusivity resemble results for
diffusion without a ratcheting mechanism, i.\,e.\,, $v=-\bareDiff f$ and $D =
\bareDiff$ since the stationary probability for a chaperone to be
bound gets smaller as $K$ grows. Therefore, fewer transitions across the
periodic boundary are rejected for large $K$. For small $K$ the movement is
affected by the chaperones and we see a plateau developing in the $f$
dependence of $v$. For small $f$ the strand is pulled to the trans side. With
decreasing $K$ the diffusion coefficient develops a minimum and a maximum. The
minimum stays roughly at the stall force while the maximum that is larger than
the bare diffusivity moves to large $f$ as $K$ decreases.}
	\label{fig:oster_diffusion_K}
\end{figure*}

\begin{equation}
D = \frac{\bareDiff f^{2} }{2 \left(K f \eul^{f} - K f - f + \eul^{f} -
1\right)^{3}} \,	G(f,K)\,,
\end{equation}
where the function $G(f,K)$ is given by the expression
\begin{align}
	G(f,K) \equiv  &(\eul^f -1)^3 [ 2 K^3 f + 6 K^2 + K ] \nonumber\\
				   &+(\eul^f -1)^2 [ -6 K^2 f + 4Kf -10K +1 ]\nonumber\\
				   &+(\eul^f -1)[ 10Kf -4f+6 ] -6f\,.
\end{align}

Figure~\ref{fig:oster_diffusion_K} shows the mean velocity and the diffusivity
for different dissociation constants $K$ as a function of the force $f$. For
$K \gg 1$ it is increasingly unlikely to find a bound chaperone that stalls the
movement. Consequently, the system behaves as if no ratcheting mechanism were
present and we find $v \approx -\bareDiff f$ and $D \approx \bareDiff $.
If $K$ is small, however, the effects of the chaperones on the diffusion get
more pronounced.

From the results for mean velocity and diffusion coefficient, it is possible to
determine how the system behaves in the limit of strong pulling force. We
find that the mean velocity diverges for $f \rightarrow \infty$, whereas the
ratio of velocity to force converges to
\begin{equation}
	\lim_{f \rightarrow \infty} \frac{v}{f} = -\bareDiff\,,
	\label{eq:mu_inf}
\end{equation}
which resembles the mobility we would have expected if no ratcheting mechanism
were present.
The same is true for the diffusion coefficient that approaches
\begin{equation}
	\lim_{f \rightarrow \infty} D = \bareDiff\,.
	\label{eq:D_inf}
\end{equation}
This means that, in the classical model, strong pulling forces negate the
effect of the ratcheting mechanism. This result is counterintuitive since one
would expect that the ratchet can stall the translocation also for large
forces and consequently there should be an impact on velocity and diffusion
coefficient similar to the effect we found in the simple model.
As it turns out, the results from Eqs.~\eqref{eq:mu_inf}
and~\eqref{eq:D_inf} are only valid if the limit $\won, \woff
\rightarrow \infty$ is performed first, which is the main assumption of the
classical model.

\section{Translocation ratchet with memory}
\label{sec:oster_model_with_memory}
\subsection{The model}
\label{sub:the_model}

The established classical model from~\cite{peskin_cellular_1993} has the
drawback that it is only valid in the limit of fast binding and unbinding
because it does not keep track of the state of the individual binding sites.
Instead, the state of a site is drawn from the stationary distribution as
needed. For finite binding rates this strategy is not justified, since the
binding sites are always empty when they enter the solution through the pore
and are therefore not in chemical equilibrium with the surrounding.

Ideally one would have to incorporate the state and dynamics of each binding
site into the model and determine whether a site can cross through the pore
back to the cis side based on the state of the specific site in front of the
pore on the trans side. For a chain of finite length this approach has been
used to determine the mean translocation time from simulations
in~\cite{liebermeister_ratcheting_2001a}. This treatment, however, is not
viable for long chains since the state space grows exponentially with the
number of binding sites taken into account.

One can, however, assume that sites far away from the membrane are in chemical
equilibrium since they did not interact with the wall for a time longer than
their equilibration time. For this reason, it is only necessary to incorporate a
finite number of sites into the model in order to obtain a satisfactory result.

Following these considerations, we propose a modified model that is defined as
follows: As in the classical case the position of the strand is represented
by a continuous variable $x$ but is augmented by the state of the first $m$
sites on the trans side, which act as a memory that we denote with $M$. As in
the case without memory, the position $x$  can take values from the interval
$[0,1)$ and evolves according to the Langevin equation~\eqref{eq:langevin}.

Each bit of the memory represents the state of one binding site, where we use 1
to represent the bound state that cannot pass through the pore and 0
corresponds to the unbound state. The leftmost bit is the site closest to the
pore. Since binding and unbinding occur randomly with the rates $\won$ and
$\woff$, each bit of the memory flips from one state to the other according to
Markovian dynamics with these rates.

The memory and position are only coupled when $x$ attempts to diffuse across
the periodic boundary. A jump from $x=0$ to $1$ means that a
binding site from the trans side moves into the pore. This process is only
possible if the site is unoccupied. So if the leftmost bit of $M$ is 1,
the jump of $x$ is rejected. If the leftmost bit is $0$, the position
variable can cross the boundary and the memory is shifted by one bit to the
left. To keep the length of the memory constant, the rightmost bit is
filled with a state drawn from the stationary distribution. We thereby assume
that this binding site is sufficiently far away from the pore that it had time
to reach chemical equilibrium.

\begin{figure}
	\includegraphics[scale=1]{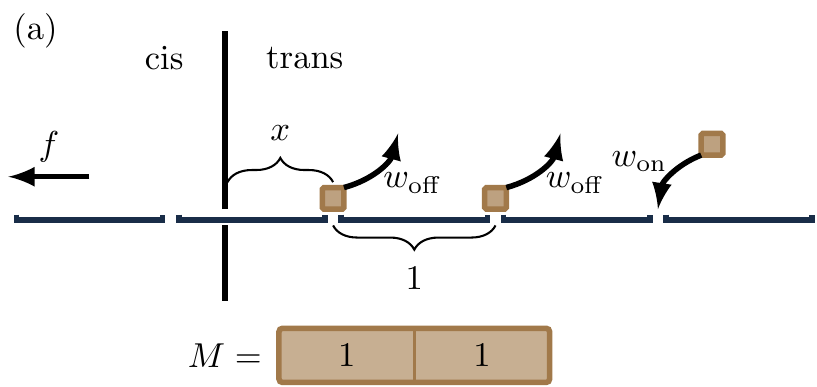}

	\vspace{2ex}

	\includegraphics[scale=1]{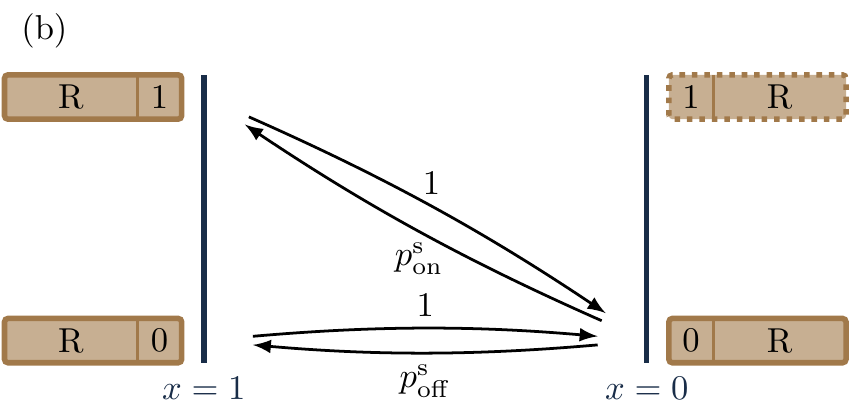}
	\caption{Model with memory. The state of the first $m$ binding
		sites after the pore on the trans side are incorporated into the
		description as it is indicated in panel (a) by a box that represents the
		state of a two-bit memory.
		Panel (b) illustrates the transition rules across the periodic boundary. The
		figure shows the state of the memory before and after the jump. When
		the position $x$ crosses the boundary from 1 to 0, the memory gets
		shifted to the right and the first bit is filled with 0. Movement
		in the opposite direction is only possible if the memory starts with
		0. In this case the memory is shifted to the left and the rightmost bit
		is filled randomly according to the stationary distribution of the
	bound or unbound state. $R$ denotes the $(m-1)$ remaining bits of the
memory that get shifted by the jump across the periodic boundary.}
	\label{fig:oster_memory_jump}
\end{figure}

A jump of $x$ in the opposite direction is always possible. If such a jump
occurs, the bits of the memory are shifted to the right and the leftmost bit is
filled with a 0 since every site is unoccupied upon entering the trans side.
The transition probabilities from one memory state to another when a jump
across the periodic boundary is attempted are depicted in
Fig.~\ref{fig:oster_memory_jump}, where the boxes represent the state
of the memory.

\subsection{Results}
\label{sub: oster_results}
For this model, the calculation of mean velocity and diffusion coefficient is much more
involved than for the previous models since we have to deal with the coupled
dynamics of both the memory and the position variable $x$. An analytical
calculation as it is done in Appendix~\ref{sec:analytical_diffusion_oster} for
the classical model is, to our knowledge, no longer feasible.
One can, however, discretize the state space and solve the systems of linear
equations that correspond to Eqs.~\eqref{eq:eigen_expansion} numerically.

\begin{figure*}
	\includegraphics[scale=1]{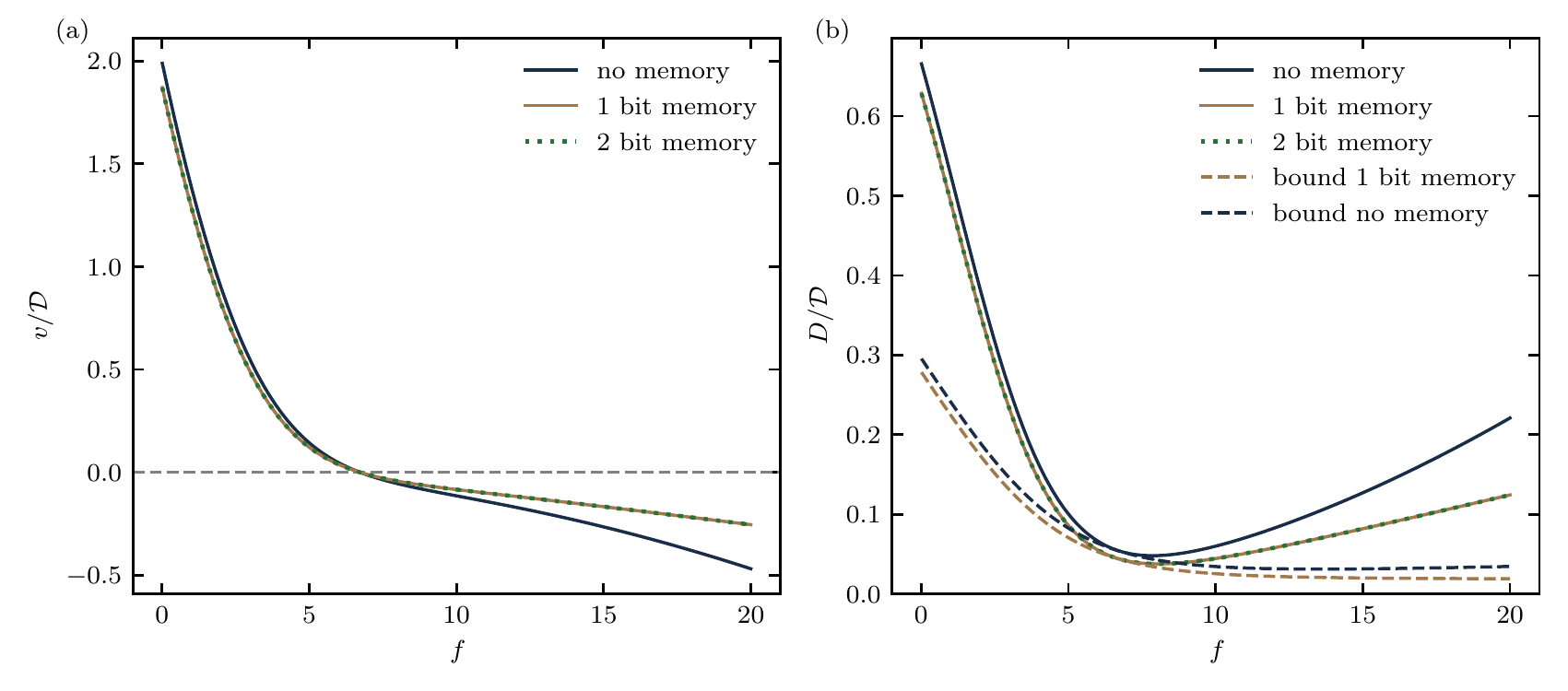}
	\caption{Comparison of velocity (a) and diffusivity (b) in the model proposed
	in~\cite{peskin_cellular_1993} and in our model with memory that
	incorporates the state of the first few binding sites. The parameters were
	chosen such that the prediction of the classical model matches the
	experimental findings from~\cite{hepp_kinetics_2016} up to rescaling of the
	axes. It is evident that the two models deviate most strongly for high
	pulling forces while the stall force is the same in both cases. The plot in
	the right panel shows the diffusion constant for both models as well as a
	lower bound following as a consequence of the thermodynamic uncertainty
	relation. Parameters: $\bareDiff = \unitfrac[1]{1}{s}$, $\won =
	\unitfrac[1000]{1}{s}$, $\woff=\unitfrac[1.2]{1}{s}$.}
	\label{fig:oster_memory_diffusion}
\end{figure*}

Figure~\ref{fig:oster_memory_diffusion} shows a comparison of the mean velocity
and the diffusion coefficient predicted by the Brownian ratchet model with and
without memory. The parameters were chosen in such a way that the results
match the experimental findings presented in~\cite{hepp_kinetics_2016} up to
scaling of the axes.

Since we no longer assume timescale separation between the diffusion and
binding process, the binding rate $\won$ and the unbinding rate $\woff$ enter
the model separately. To keep the parameters consistent
with~\cite{hepp_kinetics_2016}, where their ratio was found as $K=0.0012$, we
choose $\won=\unitfrac[1000]{1}{s}$ and $\woff=\unitfrac[1.2]{1}{s}$. The bare
diffusivity $\bareDiff$ fixes the timescale of the diffusion of the strand.
The fact that the experimental data are matched quite well by the prediction of
the classical model without memory indicates that the assumption of a timescale
separation between diffusion and (un)binding is at least approximately
valid. To reflect this fact, we choose $\bareDiff=\unitfrac[1]{1}{s}$, which is
smaller than both rates. Note that the diffusivity has the unit of a frequency
since we set the distance of two neighboring sites to unity.

It is evident that the introduction of memory into the model leads to slower
average speed of the strand both when the force is below the stall
force and the strand is pulled inside by the ratcheting mechanism and also for
forces above the stall force when the strand is pulled out. These two effects
are both caused by the memory but in a slightly different way.

Each binding site that enters the trans side is initially empty. This means
that the probability to find a bound site right in front of the pore is lower
than the equilibrium value $\pbound$. As a result the sites can diffuse back
to the cis side with a higher frequency than assumed in the model without
memory, leading in turn to smaller $v$.

With increasing $f$ and especially above the stall force this effect becomes
less and less relevant since it becomes more and more unlikely that an empty
binding site can diffuse from the cis to the trans side. Instead, we find that
the memory also decreases the mean velocity when the strand is pulled out. In
this case the difference between the two models can be explained by the fact
that in the model without memory the decision whether a site can pass through
the pore on each attempt is independent from the last attempt. If there is no
timescale separation between the binding dynamics of the chaperone molecules
and the diffusion of the strand, the outcome of successive attempts to cross
the periodic boundary should, however, be correlated. If a chaperone blocks one
attempt to diffuse outward, it will also block all attempts in the future
unless it unbinds. Introducing the memory into the model makes it possible to
capture this behavior.

The comparison of the force dependence of the diffusion coefficient presented
in Fig.~\ref{fig:oster_memory_diffusion} shows that the memory leads to
a smaller diffusivity throughout. The effect is especially pronounced for
forces well above the stall force. It is interesting to note that even the
introduction of one single bit of memory changes the behavior. The
classical model does not even keep track of the binding site right in front of
the pore, but instead draws its state from the stationary distribution every
time it is needed, which leads to differences in the mean velocity.
The figure also shows lower bounds on the
diffusion coefficient that are obtained using the thermodynamic uncertainty
relation (see also section~\ref{ssec:thermodynamics}).

\subsection{Velocity in the limit of strong driving}
\label{sub:velocity_in_the_limit_of_strong_driving}

One of the key differences between the model with and without memory is that the
predicted mean velocity differs vastly when the pulling force is large. In the
classical model, which does not keep track of the states of the binding sites,
the velocity diverges to negative infinity in the limit of strong pulling $f
\rightarrow \infty$. The numerical results for the model with memory, however,
indicate that the velocity approaches a finite value in this regime. The aim of
this section is to understand this difference in behavior.

Suppose the pulling force is large. This means that, whenever there is an
unoccupied binding site in front of the pore on the trans side, it will
immediately get pulled out. As a consequence, the movement of the strand is
determined only by the occupation pattern of the binding sites. The pulling is
stalled whenever there is an occupied site in front of the pore. Once the
blocking chaperone molecule unbinds, the strain advances suddenly to the
position of the next site that is bound and so on.
From these considerations, it is possible to calculate the mean velocity in the
limit $f \rightarrow \infty$. To do so, we need to know the average time during
which the movement is stalled by a bound site, which we call
$\tau_\text{b}$, and the average number of sites $\langle n \rangle$
the strain is advanced if it can move.

The waiting time $\tau_{\text{b}}  = 1/\woff$ is the inverse of the unbinding
rate.  The probability to advance by $n$ steps, once the movement becomes
possible, is given by
\begin{equation}
	p(n) = \pbound \left( \punbound \right)^{n-1}
	\label{eq:strong_f_jump_prob}
\end{equation}
if we assume that the state of each individual site is distributed according to
the equilibrium distribution. In the limit of strong pulling this is justified
since no empty site can enter from the cis side. From
Eq.~\eqref{eq:strong_f_jump_prob} the average number of steps is obtained as
\begin{align}
	\langle n \rangle &= \sum_{n=1}^{\infty} n \pbound \left( \punbound \right)^{n-1}
	= \frac{\pbound}{\punbound} \frac{\mathrm{d}}{\mathrm{d}a} \sum_{n=0}^{\infty}
	(a \punbound)^n \biggr|_{a=1} \nonumber \\
	&= \frac{\pbound}{\punbound} \frac{\diff}{\diff a} \left( \frac{1}{1- a \punbound}
	\right)
	\biggr|_{a=1}  =\frac{1}{\pbound} \,.
\end{align}
Therefore, the mean velocity will reach
\begin{equation}
	v_{\infty} = -\frac{\langle n \rangle}{\tau_{\mathrm{b}}} = -\frac{\won +
	\woff}{\woff/\won}
	\label{eq:v_f_inf}
\end{equation}
as $f$ goes to infinity. The minus sign is introduced because the strain is
pulled in the negative direction. For finite values of $f$ we expect that the
absolute value of the mean velocity is lower than the asymptotic value since it
takes the strain a finite amount of time to move to the next stalling position
due to friction.

\begin{figure}
	\centering
	\includegraphics{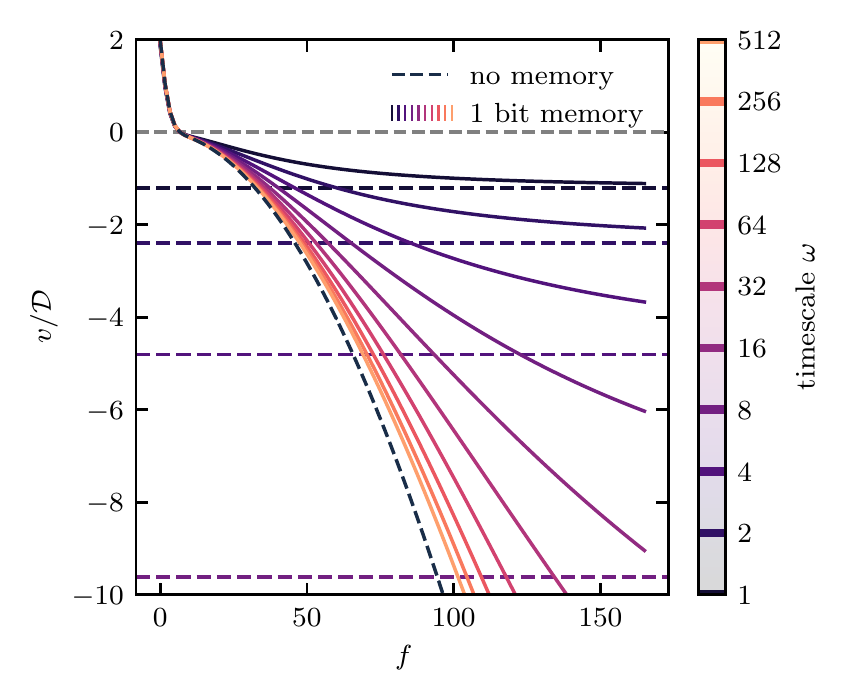}
	\caption{Comparison of the asymptotic behavior of the model with and
		without memory. Without memory the strain can get pulled out
		arbitrarily fast as the force $f$ increases. With memory the large $f$
		behavior is governed by the timescale introduced by the binding and
		unbinding rates $\won$ and $\woff$ entering Eq.~\eqref{eq:v_f_inf}.
		The asymptotic values of the velocity are indicated as dashed lines.
		Parameters: $\bareDiff = \unitfrac[1]{1}{s}$, $\won = \omega \cdot
	\unitfrac[1000]{1}{s}$, $\woff= \omega \cdot \unitfrac[1.2]{1}{s}$}
	\label{fig:oster_large_f}
\end{figure}

From Eq.~\eqref{eq:v_f_inf} it is obvious that $v_{\infty}$ diverges in the
situation assumed in the classical model ($\won, \woff \rightarrow \infty$). On
the other hand, this means that deviations of experimental data from the
predictions of the classical model can be used to infer the (un)binding rates.
This behavior is illustrated in Fig.~\ref{fig:oster_large_f}. It shows the
mean velocity for the same parameters as in
Fig.~\ref{fig:oster_memory_diffusion} over a larger range of $f$ and for
different timescales of the binding and unbinding rates. Also shown are the
asymptotic values of the velocity for each case. As the rates increase the mean
velocity approaches the result of the model without memory as
expected since the timescale separation is the key assumption of the classical
model without memory. Also, it takes larger pulling forces to reach the
asymptotic value as the rates grow.

\subsection{Minimum memory length}
\label{sub:minimum_memory_length}

The key assumption in the ratchet model with memory is that the last bit of the
memory, representing the binding site farthest away from the pore, is always in
chemical equilibrium with the surrounding. Whether this is actually the case
depends on the system parameters and especially on the length of the memory.
This section is concerned with the question of how many sites have to be modeled
explicitly to satisfy the assumption mentioned above.

\begin{figure}
	\includegraphics[scale=1]{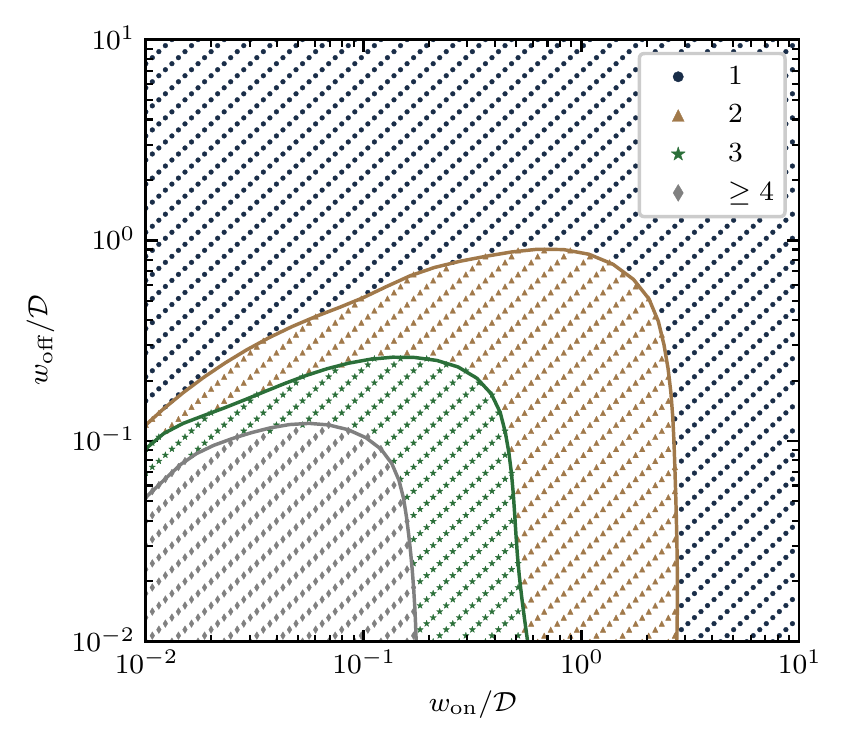}
	\caption{Minimum number of memory bits needed to accurately model the
	ratcheting mechanism as a function of the binding and unbinding rate.
	Interpolated boundaries between regions with different minimal memory
	length are added as guide to the eye.
}
	\label{fig:oster_memory_convergence}
\end{figure}

If a specific binding site is already in chemical equilibrium, so is its right
neighbor. For this reason, the prediction of the memory model should converge
with increasing memory length $m$. For the parameter values used in
Fig.~\ref{fig:oster_memory_diffusion}, one bit of memory is already
sufficient since the addition of a second bit (shown as a dotted line) does not
change the result significantly.
To calculate the minimum memory length $m_\mathrm{min}$ for a broader set of
parameters, we calculated the velocity for different (un)binding rates for
pulling forces in the range from zero to twenty. The results obtained for the
specific choice of $\bareDiff=1$ are universal, since any other choice can be realized
by rescaling in time. As a criterion for sufficient convergence, we checked
whether increasing the memory size by one bit changes the result by more than
5\% for any $f$. The resulting minimal memory length as a function of the rates
is shown in Fig.~\ref{fig:oster_memory_convergence}.

Note that we only compare variants of the model with memory of different
lengths $m$ to each other. Whether the classical model without memory is
sufficient depends strongly on the driving force, since the classical model
predicts a fundamentally different large $f$ behavior, as we have seen in
Sec.~\ref{sub:velocity_in_the_limit_of_strong_driving}. Models with memory
but different $m$ show the same asymptotics in $f$ but differ for small forces.

The plot shows that if either the binding or the unbinding rate becomes too
small, an additional bit of memory is required. The shaded areas have
approximately rectangular shape, which means that the minimum number of bits is
roughly determined by a weighted maximum of the binding and the unbinding rate.

\subsection{Affinity and stall force}
The goal of this section is to determine the affinity, i.\,e.\,, the entropy
production of one step of the Brownian ratchet with or without memory. The
entropy production is the logarithmic ratio of the probability to observe a
trajectory and the probability to observe the time reversed trajectory. Setting
the affinity to zero and solving for the driving force allows us to calculate
the stall force.

First, we specify what we mean exactly by one step of the ratchet. Suppose the
system starts at position $\xinit$ and the memory is in state $M$. After
some time it reaches the ratcheting position $x=1$ with the memory in state
$M'$. The jump across the periodic boundary changes the memory state to $M''$
according to the rules of the model. Afterwards, the system diffuses back to
position $\xinit$ without crossing the periodic boundary and the memory reverts
to $M$. Because the dynamics is Markovian and the memory and the position
evolve independently of each other, as long as the position does not try to
cross the periodic boundary, it is possible to split the affinity into
contributions associated to different sections of this trajectory.

On its way from the initial position $\xinit$ to $x=1$ the position contributes
$\mathcal{A}_{x,1} = (\xinit - 1)f$ entropy production.
The entropy production stemming from the memory during this section is
determined by the difference of bound sites.
For each bit that flips from 0 to 1, the entropy production increases by
$\ln(w_{\mathrm{on}}/w_{\mathrm{off}}) = \bindEn$.
If we use $n$ to denote the number of bound sites in $M$ and $n'$ for the bound
sites in $M'$, the entropy production in the memory reads $\mathcal{A}_{M
\rightarrow M'} = (n'-n)\bindEn$.
For the same reasons, the entropy production through the position and memory
aggregated after the jump across the boundary are $\mathcal{A}_{x,2} = -\xinit
f$ and $\mathcal{A}_{M'' \rightarrow M} = (n-n'') \bindEn$, respectively.

The jump across the periodic boundary also leads to a crucial contribution to
the entropy production since a jump in one direction is not as likely as the
jump in the backwards direction. Here, we have to distinguish between two cases:
Either the memory before the jump ends with a zero or a one. In either case the
logarithmic ratio of forward to backwards probability is
$\mathcal{A}_{\mathrm{jump,i}} = -\ln p_{i}^{\mathrm{s}}$, where $i$ is the
last bit of $M'$. Depending on $i$, the number of bound bits may be changed
through the jump. If $i=1$, the memory shift reduces the number of bound bits by
one since the first site is always unbound after the jump. For this reason, we
find in this case $n'' = n'-1$.
In the case of an unbound last site, the number of bound sites stays untouched,
leading to $n''=n'$.

Summing up, we find that the total affinity of one step is
given by
\begin{equation}
	\mathcal{A} = -f + \left\lbrace \begin{array}{ll}
		- \ln \pbound + \bindEn & \mathrm{for}\;i=1\\
		- \ln \punbound & \mathrm{for}\;i=0
	\end{array}  \right.\,.
\end{equation}
Using the dependence of the stationary distribution on the binding energy
$\bindEn$, one can easily show that both cases yield the same result
\begin{equation}
	\aff = \ln \left( \exp( \bindEn ) +1 \right) -f \equiv \freeEn -f\,,
	\label{eq:free_energy}
\end{equation}
where we identified the free energy difference $\freeEn$ associated with the
insertion of one binding site in equilibrium with the chemostat that has been
previously identified in a similar context
in~\cite{ambjornsson_chaperone-assisted_2004}.

The free energy difference introduced in~\eqref{eq:free_energy} can be split
into a term representing the average change in binding energy caused by an
additional binding site and a term representing the change in configuration
entropy
\begin{equation}
	\freeEn = \pbound \bindEn -\punbound\ln \punbound  -\pbound\ln \pbound\,.
\end{equation}
This form of the free energy difference makes it especially clear that the
strand is pulled in, even if the binding energy of the chaperones vanishes, due
to the increase in entropy in the memory. Conversely, to pull out one binding
site one has to delete information from the memory, which, according to
Landauer's principle, comes at an energetic cost. In the case $\bindEn = 0$ we
indeed have $\freeEn = \ln 2$ as it is to be expected.
So in some sense a Brownian ratchet can be seen as a stochastical information
processing machine that can erase from or write bits to a memory (see also
\cite{mandal_work_2012, barato_autonomous_2013} for further examples of such
devices).

The stall force follows trivially from the condition that the affinity
vanishes, leading to
\begin{equation}
	f_0 = \freeEn\,.
\end{equation}
Incidentally, this is the same result as already obtained
in~\cite{peskin_cellular_1993} for the classical model without memory.

\subsection{Thermodynamics and efficiency}
\label{ssec:thermodynamics}

\begin{figure}
	\centering
	\includegraphics[scale=1]{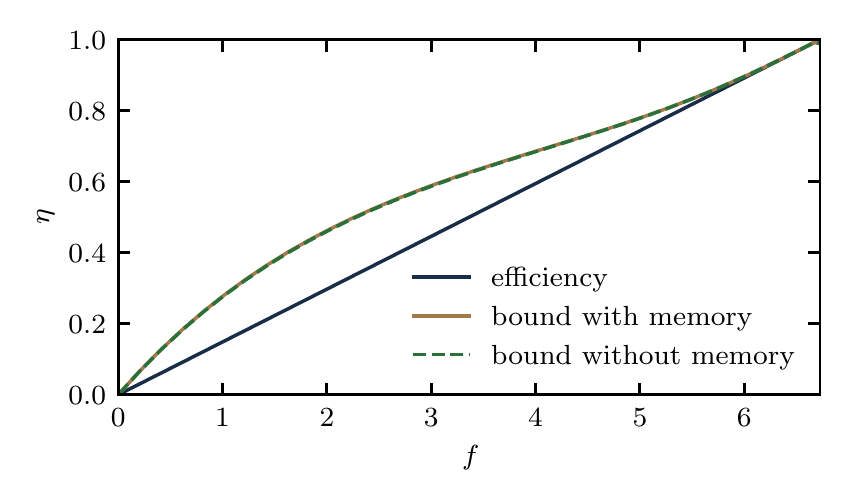}
	\caption{Efficiency of the Brownian ratchet and upper bound derived from
	the classical model with and without memory. The bound becomes tight for forces
	close to the stall force. The difference between the models with and without
	memory is barely visible since they only deviate significantly for forces
	well above the stall force, where the motor does not perform work.}
	\label{fig:efficiency}
\end{figure}

The analysis of the previous section showed that the Brownian ratchet can be
seen as a kind of molecular motor that uses the free energy $\freeEn$ gained by
providing an additional binding site  to perform a certain amount of work $\Delta w
= f$. It is therefore natural to identify the efficiency $\eta$ of the Brownian
ratchet as the ratio of the performed work to the maximum possible amount,
i.\,e.\,,
\begin{equation}
	\eta \equiv \frac{\Delta w}{\freeEn} = \frac{1}{1+\aff/f } \leq 1 \,.
\end{equation}

The thermodynamic uncertainty relation that was recently conjectured
\cite{barato_thermodynamic_2015, pietzonka_universal_2016} based
on extensive numerical evidence and later proven
\cite{gingrich_dissipation_2016, horowitz_proof_2017a} implies a boundary on the
efficiency that only relies on the observable quantities $v$, $D$ and
$f$~\cite{pietzonka_universal_2016a}.

The uncertainty relation states that precision, i.\,e.\,, low diffusion $D$
comes at the cost of high dissipation.
For the system at hand, it can be stated in the form
\begin{equation}
	\frac{D}{v} \aff \geq 1\,.
	\label{eq:uncertainty}
\end{equation}
This relation has a number of consequences regarding the force dependence of
mean velocity and diffusion coefficient.

Since the affinity of a step of the ratchet is known [see
Eq.~\eqref{eq:free_energy}], it is possible to derive a lower bound on the
diffusion coefficient by solving Eq.~\eqref{eq:uncertainty} for $D$.
The resulting bounds are shown as dashed lines in
Fig.~\ref{fig:oster_memory_diffusion}.
Since the bound becomes tight in the linear response regime and close to the
stall force, it allows also for the calculation of the diffusion coefficient at
stall force:
\begin{equation}
	D_{\mathrm{stall}} = \lim_{f \rightarrow f_{0} }\frac{v(f)}{\freeEn
		- f} = - \left. \frac{\partial v(f)}{\partial f} \right|_{f=f_{0} } \,.
\end{equation}

In terms of the efficiency of the ratchet the uncertainty relation is expressed
in the inequality
\begin{equation}
	\eta \leq \frac{1}{1+\frac{v}{Df}}\,.
\end{equation}
The efficiency as well as the bounds derived from the model with or without
memory are shown in Fig.~\ref{fig:efficiency}. As expected, the bound becomes
tight in the linear response regime and close to the stall force.

\section{Summary}
\label{sec:conclusion}
We have analyzed the force dependence of the diffusion coefficient for three
different models of Brownian ratchets with varying degree of complexity.
For the first model that describes the ratcheting mechanism in terms of an
asymmetric random walk with an alternating set of rates one can obtain the
scaled cumulant generating function in closed form. We derived a bound for the
Fano factor at vanishing pulling force that is stronger than the bound obtained
from the thermodynamic uncertainty relation. For this model we identified
different regimes in the parameter space. Depending on the regime the force
dependence of the diffusion coefficient can develop a local minimum.

For a classical model for the Brownian ratchet that was first introduced
in~\cite{peskin_cellular_1993} and has been found to explain
DNA uptake by Neisseria gonorrhoeae bacteria~\cite{hepp_kinetics_2016}, we
derived an expression for the diffusion coefficient. This new result could in
the future be used for comparison with experimental data.

The classical model is based on the assumption that the binding and unbinding
of the chaperones are fast compared to the diffusion process. We found that the
large $f$ behavior does not match the expectation of a ratcheted process, due
to this timescale separation. In order to overcome the limitations of the
classical model we introduced a memory into the model that allowed us to
numerically calculate predictions for the force dependence of mean velocity and
diffusion coefficient in the case where there is no clear timescale
separation. Here we found that the mean velocity approaches a finite value as
$f$ increases, as one would expect from naive considerations.
We could identify the stall force for the model with memory and found that it
matches the result of the model without memory. We also analyzed how many bits
of memory are needed for the results to converge to their respective values if
an infinite memory were used. We found, in accordance with our
expectations, that the more memory is needed the slower the (un)binding process
is. Conversely, we confirmed that in the limit of infinitely fast (un)binding no
memory is needed at all.

We identified the thermodynamic efficiency of the ratchet and illustrated how
one can use the measurable quantities $v$, $D$, and $f$ to bound the efficiency
from above. This bound is a consequence of the thermodynamic uncertainty
relation and becomes tight for forces around the stall force.

\section*{Acknowledgments}
We thank Berenike Maier for stimulating discussions and Paolo De\mbox{
}Los\mbox{ }Rios for
valuable feedback on the paper.
\appendix

\section{Systematic scheme for the calculation of cumulants of integrated
currents}
\label{sec:diffusion_coefficient}

In this appendix, we recall a general formalism that can be used to determine the
diffusion coefficient in Markov processes like the models for Brownian ratchets
considered in this paper~\cite{koza_general_1999}.
It is based on methods commonly used in the field of large
deviations~\cite{touchette_large_2009}.
The mean velocity, the diffusion coefficient and in fact all rescaled
cumulants of the distribution of the traveled distance are encoded in the
rescaled cumulant generating function
\begin{equation}
	\alpha(\lambda) \equiv \lim_{t \rightarrow \infty}  \frac{1}{t}
		 \ln \left\langle \eul^{\lambda \Delta x}  \right\rangle
\end{equation}
as its derivatives at $\lambda=0$. Expanding the generating function in a
Maclaurin series up to second order yields
\begin{equation}
	\alpha(\lambda) = v \lambda + D \lambda^2 + \bigO(\lambda^3)\,.
\end{equation}
The generating function can be obtained as the Perron-Frobenius eigenvalue of
the so called tilted operator that generates the evolution of the moment
generating function conditioned on the final state of the trajectory
\begin{equation}
	g(\lambda, x, t) \equiv	\left\langle \eul^{\lambda \Delta x}
	\right\rangle_{x} \,,
\end{equation}
where we use the notation $	\left\langle \cdot \right\rangle_{x}$ to denote an
average over all trajectories that end in the state $x$. The time evolution of
this quantity is given by
\begin{equation}
	\partial_{t} g(\lambda, x, t) = \tiltOp(\lambda) g(\lambda, x, t)\,,
\end{equation}
where the tilted operator $\tiltOp(\lambda)$ is a differential operator
acting on the state argument of $g(\lambda, x, t)$. In the case of $\lambda=0$
the tilted operator is identical to the Fokker-Planck or master operator
$\tiltOp_{0}$ generating the time evolution of the probability distribution.
Typically, the eigenvalue problem leading to the generating function cannot be
obtained in closed form. The cumulants, however, can be calculated iteratively
by expanding the eigenvalue equation in orders of $\lambda$.  To this end, we
insert the expansions
\begin{subequations}
	\begin{align}
		\tiltOp(\lambda) &= \tiltOp_{0} + \tiltOp_{1}\lambda +
		\tiltOp_{2}\lambda^2 + \bigO(\lambda^3) \\
		\vec{q}(\lambda) &= \vecStatDist + \vec{q}_1 \lambda + \vec{q}_{2}
		\lambda^2 + \bigO(\lambda^3)
	\end{align}
\end{subequations}
into the eigenvalue equation $\tiltOp(\lambda) \vec{q}(\lambda) = \alpha(\lambda)
\vec{q}(\lambda)$, sort by orders of $\lambda$, and arrive at
\begin{subequations}
	\begin{align}
		\tiltOp_{0} \vecStatDist &= 0\,, \\
		\tiltOp_{0} \vec{q}_{1} + \tiltOp_{1}\vecStatDist &= v \vecStatDist
		\label{eq:order_one}\,,  \\
		\tiltOp_{1}\vec{q}_{1} + \tiltOp_{0}\vec{q}_{2} + \tiltOp_{2} \vecStatDist
		&= D \vecStatDist + v \vec{q}_{1}\,. \label{eq:order_two}
	\end{align}
	\label{eq:eigen_expansion}
\end{subequations}
Since the constant function denoted by the vector $\bra{1}$ is a left
eigenvector to the Markov operator $\tiltOp_{0}$  with corresponding eigenvalue
zero, the solutions of these linear equations are not unique.  They become unique,
if we additionally demand that they are normalized, i.\,e.\,, $\braket{1|\vecStatDist}
=  1$ and $ \braket{1| \vec{q}_{1}} = \braket{1|\vec{q}_{2}} = 0$, where
$\braket{\cdot | \cdot}$ denotes the standard scalar product.
To calculate velocity and diffusion coefficient, we multiply
equations~\eqref{eq:order_one} and~\eqref{eq:order_two} by $\bra{1}$ from the
left and use $\bra{1} \tiltOp_{0} = 0$, leading to
\begin{subequations}
	\begin{align}
		v &= \Braket{1 | \tiltOp_{1} | \vecStatDist}
		\label{eq:mean_vel} \\
		D &= \Braket{1 | \tiltOp_{1} | \vec{q}_{1}} + \Braket{1 | \tiltOp_{2} |
			\vecStatDist}
		\label{eq:diffusion}
	\end{align}
	\label{eq:cumulants}
\end{subequations}

Numerically this set of equations can be solved by projection of the vectors
and operators in an appropriate set of basis vectors and solving the
corresponding set of linear equations.

\section{Calculation of the diffusion coefficient in the classical model}
\label{sec:analytical_diffusion_oster}

In this appendix, we derive a closed-form expression for the diffusion
coefficient in the classical model for a Brownian ratchet without memory.
To this end, we aim to solve Eqs.~\eqref{eq:cumulants} that allow us to
calculate the diffusion coefficient using the tilted operator
$\mathcal{L}(\lambda)$.

The boundary condition used in~\cite{peskin_cellular_1993} states that
transitions from the left border of the region at $x=0$ to the right border at
$x=1$ are rejected with probability $\pbound$, while transitions in the
opposite direction are always allowed. This rather unorthodox form of a
periodic boundary condition cannot be intuitively expressed in terms of a
continuous description of the state space. To circumvent this issue, we
discretize the state space, perform all necessary calculations, and take the
limit of infinitely small discretization at the end.

\subsection{Discretization}
\label{sub:discretization}

We split the state space into $N$ discrete points separated by the distance
$h=1/N$. Therefore each continuous function $\psi(x)$ is approximated by the
$N$-dimensional vector $\psi_{i} \equiv \psi(x_{i})$ with $x_{i} = i h$.
In this picture, the time evolution of the probability distribution $p_{i}$ is
given by the master equation
\begin{multline}
	\partial_{t}p_{i} = (\tiltOp_0 p)_i \equiv p_{i-1} \left(-\frac{\bareDiff f}{2h}  +
		\frac{\bareDiff }{h^2} \right) + p_{i+1} \left(\frac{\bareDiff f}{2 h} +
	\frac{\bareDiff }{h^2}\right) \\
	- p_0 \left( \frac{2 \bareDiff }{h^2} \right) \quad i \not\in \{1,N\}
\end{multline}
that approximates the Fokker-Planck equation inside the interval $(0,1)$. Due
to the nature of the boundary conditions, the states at the boundary of the
interval need special treatment.

Because jumps from $x=0$ to $1$ are rejected with probability $\pbound$, the
rate corresponding to jumps in this direction is diminished, leading to
\begin{multline}
({L}_0 p)_N = p_{N-1} \left(-\frac{\bareDiff f}{2h}  + \frac{\bareDiff }{h^2}
\right) \\
+ p_{1} \left(1-\pbound \right) \left(\frac{\bareDiff f}{2 h} +
\frac{\bareDiff }{h^2}\right)
+  p_N \left(- \frac{2 \bareDiff }{h^2} \right)\,.
\label{eq:fp_upper_boundary}
\end{multline}

Consequently, the exit rate of state $i=1$ is decreased by the same amount and
we find
\begin{multline}
	(\mathcal{L}_0 p)_1 = p_{N} \left(-\frac{\mathcal{D}f}{2h}  +
		\frac{\mathcal{D}}{h^2} \right) + p_{2} \left(\frac{\mathcal{D}f}{2 h} +
	\frac{\mathcal{D}}{h^2}\right) \\
-  p_1 \left( (2-\pbound) \frac{ \mathcal{D}}{h^2} -\pbound
		\frac{\mathcal{D}f}{2h}
		\right) \,.
\end{multline}

In the discrete picture, we also have to replace integrals by their Riemann sum
approximation. So, for example, the normalization condition of the stationary
distribution becomes
\begin{equation}
	\int_{0}^{1} \statDist(x)\, \mathrm{d}x \approx h \sum_{i=1}^{N}
\statDist_{i}   =1\,.
\end{equation}

\subsection{Tilted operator}
\label{sub:tilted_operator}
The first step in our derivation is the identification of the tilted operator
$\mathcal{L}(\lambda)$, or to be more precise, the Taylor expansion of this
operator up to second order in $\lambda$.

While the boundary conditions prevent us from using known results for the
tilted operator in continuous state space, we can use the general result
\begin{equation}
	\mathcal{L}(\lambda)_{i,j}  = \mathcal{L}_{0,i,j}  \exp(d_{j,i} \lambda)
\end{equation}
for a discrete state space, where the distance matrix is given by $d_{i,j} = h
(\delta_{i,j+1} - \delta_{i,j-1})$.

Expanding this expression into a Taylor series, we find
\begin{equation}
	\tiltOp_{1,i,j} =  \tiltOp_{0,i,j} d_{j,i}
\end{equation}
and
\begin{equation}
	\tiltOp_{2,i,j} = \tiltOp_{0,i,j} d_{j,i}^{2}/2 \,.
\end{equation}

Now we discuss how these operators act on a test function $\psi(x)$ in the
continuous limit.

\subsubsection{First order}
\label{ssub:first_order} %
For the first order coefficient, we arrive at
\begin{multline}
	(\mathcal{L}_{1} \psi)_i = \psi_{i-1} \left(-\frac{\mathcal{D}f}{2}  +
		\frac{\mathcal{D}}{h} \right) - \psi_{i+1} \left(\frac{\mathcal{D}f}{2}
	+ \frac{\mathcal{D}}{h}\right)\\
	= - \mathcal{D}f \frac{\psi_{i+1} + \psi_{i-1}}{2} - \mathcal{D}
	\frac{\psi_{i+1} - \psi_{i-1}}{h}\,,
\end{multline}
which converges in the limit $h \rightarrow 0$ to the known result for a
continuous state space that is given by
\begin{equation}
	\mathcal{L}_1 = -\mathcal{D}(f+ 2\partial_x)\,.
\end{equation}

At the boundary the situation is, again, more subtle, since the rates are
modified and we also have to take into account that the function the operator
is acting on may jump across the periodic boundary.
Here, we find the relations
\begin{equation}
	(\mathcal{L}_1 \psi)_1 = -\mathcal{D}f \frac{\psi_N + \psi_1}{2} +
	\frac{\mathcal{D}}{h} \left(\psi_N - \psi_1\right)
\end{equation}
at the lower bound of the interval and
\begin{multline}
	(\mathcal{L}_1 \psi)_N = - \frac{\mathcal{D} f}{2} \left[
		\left(1-\pbound \right) \psi_1 + \psi_N \right] \\
		+
		\frac{\mathcal{D}}{h} \left[-\left(1-\pbound \right) \psi_1 +
		\psi_N \right]
\end{multline}
at the upper boundary. Both expressions contain terms that may diverge if we
take the continuous limit. Note that if $\psi(x)$ obeys the boundary condition
$\psi(1)=(1-\pbound)\psi(0)$ that holds for the stationary
distribution, the value at the upper bound stays finite while the value at the
lower bound diverges.

The apparent problem of possibly diverging values of $\tiltOp_{1} \psi$ can be
resolved by the fact that they are only ever needed inside of integrals. Since
the divergence is of order $1/h$, the result of the integral converges if
we let $h$ go to zero. However, additional terms introduced by the divergences
at the boundaries have to be taken into account.

Combining the action of the operator $\tiltOp_{1}$ inside the interval and on
its boundaries, we obtain the integration rule
\begin{multline}
	\int_0^1 \mathcal{L}_1 \psi \, \mathrm{d}x = -\int_0^1 \mathcal{D}(f \psi +2
	\psi') \, \mathrm{d}x
	+ 2 \mathcal{D} \left[\psi(1)-\psi(0) \right] \\
	+ \pbound
	\mathcal{D} \psi(0) = -\mathcal{D}f \int_0^1 \psi \, \mathrm{d}x +
	\mathcal{D}\pbound \psi(0)
	\label{eq:int_rule}
\end{multline}
that will become useful for the evaluation of the mean velocity and the
diffusion coefficient according to Eqs.~\eqref{eq:mean_vel}
and~\eqref{eq:diffusion}, respectively.

\subsubsection{Second order}
\label{ssub:second_order}
For the second order coefficient of the expansion of the tilted operator, we
perform calculations along the same line as for the first order and find that
the discretization
\begin{equation}
	(\mathcal{L}_2 \psi)_i =\frac{1}{2} \left[ \psi_{i-1}
	\left(-\frac{\mathcal{D}f}{2} h  + \mathcal{D} \right) + \psi_{i+1}
\left(\frac{\mathcal{D}f}{2}h + \mathcal{D}\right) \right]
\end{equation}
converges in the continuous limit.
Inside the interval, where the test function is continuous, the operator acts
as a scalar, allowing us to identify
\begin{equation}
	\tiltOp_{2} = \mathcal{D} \,.
\end{equation}
While this is not the case at the boundaries, it is obvious that the values of
$(\tiltOp_{2} \psi)_{1}$ and $(\tiltOp_{2} \psi)_{N}$ stay finite in the limit
$h \rightarrow 0$ even if the test function jumps. Since $\tiltOp_{2} \psi$
also appears only inside integrals, these deviations do not affect the result.

\subsection{Calculation of the cumulants}
\label{sub:calculation_of_the_cumulants}
Now that we know how the operators $\tiltOp_{i}$ act, we are able to
iteratively calculate the scaled cumulants like mean velocity and diffusivity.
Plugging the integration rule \eqref{eq:int_rule} into
Eqs.~\eqref{eq:mean_vel} and \eqref{eq:diffusion}, we find
\begin{equation}
	v = \int_0^1 \mathcal{L}_1 \statDist(x) \, \mathrm{d}x = - \mathcal{D}f
	+\mathcal{D} \pbound \statDist(0)
	\label{eq:mean_vel_boundary}
\end{equation}
and
\begin{align}
	D &= \int_0^1 \mathcal{L}_1 q_1(x) \, \mathrm{d}x + \int_0^1 \mathcal{L}_2
	\statDist(x) \, \mathrm{d}x \nonumber \\
	&= \mathcal{D} \left[1+\pbound q_1(0)\right]\,.
	\label{eq:diffusion_boundary}
\end{align}

The last step remaining is the calculation of the stationary distribution and
the first order correction of the eigenfunction of the tilted operator
$q_{1}(x)$.

\subsubsection{Stationary distribution}
\label{ssub:stationary_distribution}

The stationary distribution is the solution of the stationary Fokker-Planck
equation
\begin{equation}
	\tiltOp_{0} \statDist(x) = \mathcal{D}\left[ f \partial_{x} +
	\partial_{x}^{2}  \right] \statDist(x) =  0
	\label{eq:fp_equation}
\end{equation}
obeying the boundary condition $\statDist(1)=(1-\pbound)
\statDist(0)$. This boundary condition was
derived~\cite{peskin_cellular_1993}. It also emerges in a natural way if one
considers the discretized Fokker-Planck equation at the boundaries. In order
for Eq.~\eqref{eq:fp_upper_boundary} to converge to zero, the factors in
front of the jump rates have to be equal, which leads to the aforementioned
boundary condition.

The differential Eq.~\eqref{eq:fp_equation} has the general solution
\begin{equation}
	\statDist(x) = c_{1} \exp(-f x)  + c_{2}
\end{equation}
with the two constants $c_1$ and $c_{2}$ that are determined by the boundary
condition and the normalization of the stationary distribution. Solving the
corresponding set of linear equation yields
\begin{subequations}
\begin{align}
	c_{1} &=  \frac{f e^{f}}{K f e^{f} - K f - f + e^{f} - 1} \quad \text{and}\\
	c_{2} &= \frac{f \left(K e^{f} - K - 1\right)}{K f e^{f} - K f - f + e^{f} -
	1}\,,
\end{align}
\end{subequations}
where we used the relation $\pbound=1/(1+K)$ to express the stationary
distribution of a bound site in terms of the dissociation constant $K =
\woff/\won = \exp(-\bindEn)$.

The mean velocity follows directly from Eq.~\eqref{eq:mean_vel_boundary} and is
given by
\begin{equation}
	v = -\mathcal{D}f + \mathcal{D} \pbound \left( c_{1} + c_{2}
	\right)\,.
\end{equation}
After inserting the constants into this equation this result matches the one of
Peskin et\,al.~\cite{peskin_cellular_1993}, found in their original
publication by slightly different means.

\subsubsection{First order correction}
\label{ssub:first_order_correction}

In the next step, we calculate the first order correction to the eigenfunction
of the tilted operator that is given by the solution of the inhomogeneous
differential equation
\begin{equation}
	\tiltOp_{0} q_{1}(x) = b(x)\,
	\label{eq:q1_equation}
\end{equation}
where the inhomogeneity is given by
\begin{equation}
	b(x) \equiv v \statDist(x) - \tiltOp_{1}p_{\mathrm{s}}(x) = A
	\exp(-fx) + B + C \delta(0)\,,
	\label{eq:inhomog}
\end{equation}
with $C=-A[1-\exp(-f)]/f - B$.
The constants $A$ and $B$ are related to the constants appearing in the
stationary distribution by the relations
\begin{subequations}
\begin{align}
	A &\equiv \bareDiff[ \pbound c_{1}(c_{1}+c_{2}) - 2f c_{1}] \quad \text{and}     \\
	B &\equiv \bareDiff[\pbound c_{2}(c_{1}+c_{2}) ]\,.
\end{align}
\end{subequations}
As explained earlier, the boundary conditions used in the model lead to a
deltalike divergence stemming from the application of the operator
$\tiltOp_{1}$ to the stationary distribution. This divergence does not need to
be taken into account when solving the differential equation. It merely assures
that $b(x)$ is zero on average.  If this were not the case, one could
construct a contradiction by integrating both sides of
Eq.~\eqref{eq:q1_equation} over the whole state space. The left hand side
vanishes, since conservation of probability demands that a constant function is
a left eigenfunction of the Fokker-Planck operator with eigenvalue zero.
Consequently, the right hand side also has to vanish and $b(x)$ must be zero on
average.

The fact that the divergence is of no consequence to the solution of the
differential equation can be shown most easily in the discretized picture,
where the divergence appears in one single equation of the set of linear
equations given by
\begin{equation}
	\sum_{j} \tiltOp_{0,i,j} q_{1,j} = b_{i}\,.
\end{equation}
Since all columns of this system of equations sum up to zero, we are free to
choose any one of the equations and replace it by the condition that the
elements of $q_{1,i}$ have to sum up to zero. If we chose the equation
containing the deltalike divergence, the remaining set of equations converges
to Eq.~\eqref{eq:q1_equation} without the delta function.

The solution of this equation can be obtained using the ansatz
\begin{equation}
	q_{1}(x) = a x \exp(-fx) + b x + c \exp(-fx) +d \,.
\end{equation}
The constants have to be chosen such that Eq.~\eqref{eq:q1_equation} is
satisfied, leading to
\begin{subequations}
\begin{align}
	a &= -\frac{A}{\mathcal{D}f} \quad \text{and} \\
	b &= \frac{B}{\mathcal{D}f} \,.
\end{align}
\end{subequations}
The two remaining constants are fixed by the condition $\int_{0}^{1} q_{1}(x)
\, \mathrm{d}x = 0$ and the boundary condition $q_{1}(1) =
(1-\pbound)q_{1}(0)$. That the latter must hold can be shown, as in the
case of the stationary distribution, by considering the action of the
discretized Fokker-Planck operator at the boundary.

\subsubsection{Diffusion coefficient}
\label{ssub:diffusion_coefficient}
Now that the first order correction $q_{1}(x)$ is known, the diffusion
coefficient follows according to Eq.~\eqref{eq:diffusion_boundary} as
\begin{equation}
	D = \mathcal{D} \left[ 1 + \pbound \left( c+d \right)  \right]\,.
\end{equation}
By expressing the constants $c$ and $d$ trough the original parameters of the
model, one arrives at the final result for the diffusion coefficient
\begin{equation}
D = \frac{\mathcal{D} f^{2} }{2 \left(K f e^{f} - K f - f + e^{f} -
1\right)^{3}} \,	G(f,K)\,,
\end{equation}
with
\begin{align}
	G(f,K) \equiv  &(\eul^f -1)^3 [ 2 K^3 f + 6 K^2 + K ] \nonumber\\
				   &+(\eul^f -1)^2 [ -6 K^2 f + 4Kf -10K +1 ]\nonumber\\
				   &+(\eul^f -1)[ 10Kf -4f+6 ] -6f\,.
\end{align}

\vspace{0.5em}
\subsubsection{Higher order cumulants}
\label{ssub:higher_order_cumulants}
The calculation of the mean velocity and the diffusion coefficient presented
in the previous sections show how, in principle, all cumulants can be
obtained. Since the Fokker-Planck operator only contains derivatives up to
second order, all Taylor coefficients of the tilted operator $\tiltOp_{i}$
vanish for $i>2$. The problem of finding the $n$-th order cumulant
$\mathcal{C}_{n}$ essentially reduces to the solution of the differential
equation
\begin{equation}
	\tiltOp_{0} q_{n}(x) = b_{n}(x)
	\label{eq:high_order_q}
\end{equation}
where the inhomogeneity $b_{n}(x)$ is given by
\begin{equation}
	b_{n}(x) \equiv - \tiltOp_{1} q_{n-1}(x) - \bareDiff q_{n-2}(x) +
	\sum_{\ell=1}^{n} \frac{\cum_{\ell} }{\ell !} q_{n-\ell}(x)
\end{equation}
and we use the identification $q_{0}(x) = \statDist(x)$.
The higher order rescaled cumulants themselves are related to the expansion
coefficients of the eigenfunction by
\begin{equation}
	\cum_{n}  = n!  \, \bareDiff \pbound q_{n-1}(0) \quad \text{for }
	n>2\,.
\end{equation}

It can be shown by induction that $q_{n}$ has the form
\begin{equation}
	q_{n}(x) = \mathcal{Q}_{1,n}(x) \exp(-fx)+ \mathcal{Q}_{2,n}(x)\,,
\end{equation}
where $\mathcal{Q}_{1,n}$ and $\mathcal{Q}_{2,n}$ are two polynomials of order
$n$. With this ansatz for $q_{n}$,
Eq.~\eqref{eq:high_order_q} reduces to $2n$ linear equations for the
coefficients of these polynomials. Two additional equations are given by the
normalization condition $\int_{0}^{1} q_{n}(x) \,\mathrm{d}x = 0$ and the
modified periodic boundary condition.

\bibliographystyle{bibgen.bst}
\bibliography{literatur.bib}
\end{document}